\documentclass[twocolumn,showpacs,preprintnumbers,amsmath,amssymb,groupedaddress,pre]{revtex4}

\usepackage{graphicx}
\usepackage{dcolumn}
\usepackage{bm}

\newcommand{\unit}[1]{\,\mathrm{#1}}
\newcommand{\elem}[2]{${}^{#1} \mathrm{#2}$}

\begin{document}


\title{Scaling of hysteresis loops at phase transitions into a quasiabsorbing state}


\author{Kazumasa A. Takeuchi}
\email{kazumasa@daisy.phys.s.u-tokyo.ac.jp}
\affiliation{
Department of Physics, The University of Tokyo, 7-3-1
  Hongo, Bunkyo-ku, Tokyo, 113-0033, Japan
}%
\affiliation{Service de Physique de l'\'Etat Condens\'e, CEA-Saclay, 91191 Gif-sur-Yvette, France}


\date{\today}

\begin{abstract}
Models undergoing a phase transition to an absorbing state 
weakly broken by the addition of a very low spontaneous nucleation rate 
are shown to exhibit hysteresis loops 
whose width $\Delta\lambda$ depends algebraically on the ramp rate $r$.
Analytical arguments and numerical simulations show that
$\Delta\lambda \sim r^{\kappa}$
 with $\kappa = 1/(\beta'+1)$, where $\beta'$ is the critical exponent
governing the survival probability of a seed near threshold.
These results explain similar hysteresis scaling
 observed before in liquid crystal convection experiments.
This phenomenon is conjectured to occur 
 in a variety of other experimental systems.
\end{abstract}


\maketitle

Directed percolation (DP) is an archetypical model
 of phase transitions into an absorbing state,
 i.e. a state from which a system can never escape.
A vast literature of theoretical and numerical studies
 has enlarged the range of phenomena in the DP universality class
 \cite{Hinrichsen-AdvPhys2000},
 refining conditions for this prominent critical behavior,
 known as DP conjecture
 \cite{Janssen-ZPhysB1981,Grassberger-ZPhysB1982,Hinrichsen-AdvPhys2000}.
Experimentally, the author and coworkers recently found
 that electrohydrodynamic convection of nematic liquid crystal
 shows the scaling behavior of DP at the transition
 between two turbulent states (DSM1-DSM2)
 \cite{Takeuchi_etal-PRL2007}.
Applying voltages $V$ closely above the threshold,
spatiotemporal intermittency (STI) occurs, in which
DSM2 patches move around in a DSM1 background.
As conjectured early by Pomeau \cite{Pomeau-PhysD1986},
this STI was unambiguously mapped onto DP
 with DSM1 playing the role of the absorbing state.
This constituted a clear experimental realization
 of a DP-class absorbing phase transition.

On the other hand, Kai \textit{et al.} reported in 1989 hysteresis 
phenomena around this DSM1-DSM2 transition \cite{Kai_etal-PRL1990}.
Measuring the global light transmittance through the sample,
 increasing or decreasing the applied voltage $V$ at a rate $r$,
 they found hysteresis loops of width $\Delta V$ scaling roughly like
 $\Delta V \sim r^{\kappa}$ with $\kappa\approx 0.5$-$0.6$
 \cite{Kai_etal-PRL1990,Kai-PC1}.
In particular, these loops disappear in the small-$r$ limit,
 and it has been discussed whether the transition corresponds
 to a supercritical bifurcation or a subcritical one.
This is in apparent contradiction with DSM1 being an absorbing state,
 since then one expects infinitely wide hysteresis loops. 
It is shown here that
 the scaling of hysteresis loops is in fact in full agreement 
with the DP framework in which the DSM1 state is only quasi-absorbing,
 i.e. with the existence of a small residual
 probability for spontaneous nucleation of DSM2 patches either 
 in the bulk or at the boundaries.

As a first illustration, a probabilistic cellular automaton (PCA)
version of the contact process (CP)
 \cite{Harris-AnnProb1974,Hinrichsen-AdvPhys2000} is introduced, 
in which an extra, small probability $h$ 
 to create an active site spontaneously anywhere is added.
Consider a two-dimensional (2D) square lattice of size $L \times L$
 and assign a variable
$s_{i,j}$ to each lattice point,
encoding its local state,
 either inactive (absorbing, $s_{i,j}=0$) or active ($s_{i,j}=1$).
Indices $i$ and $j$ denote Cartesian coordinates.
The time evolution is as follows:
 randomly choose one site and stochastically flip it with probabilities
\begin{align}
 &p_{i,j}(0 \to 1) = \frac{p_1}{4}(s_{i-1,j}+s_{i+1,j}+s_{i,j-1}+s_{i,j+1}) + h,  \notag \\
 &p_{i,j}(1 \to 0) = p_2,  \label{eq:CP2D}
\end{align}
 where $p_1 = \lambda/(\lambda+1)$ and $p_2 = 1/(\lambda+1)$.
The two terms in the first equation account for contamination by neighbors
 and spontaneous nucleation of active sites, respectively.
Periodic boundary conditions $s_{i,j}=s_{i+L,j}=s_{i,j+L}$ are used throughout,
 and a time step (or Monte Carlo step, MCS)
 consists of $L^2$ flipping attempts.
The $h=0$ case is known as the PCA version of the original (2+1)D CP,
 which shows a DP-class transition at $\lambda_{\rm c}=1.64877(3)$
 \cite{Dickman-PRE1999}
 (the number in parentheses denotes the uncertainty in the last figure).
In the present study,
 $L=256$ and $h' \equiv hL^2 = 10^{-2}$.
Although, strictly speaking, even rare nucleation events wipe out
 the absorbing phase transition,
 in practice a significantly low nucleation rate allows us to observe
 the underlying critical behavior as we shall see in this study.
The nucleation rate $h$ theoretically corresponds to an external field
 \cite{Luebeck-IJMB2004},
 so a weak-field case is dealt with here.

\begin{figure*}[t]
 \begin{center}
  \includegraphics[clip]{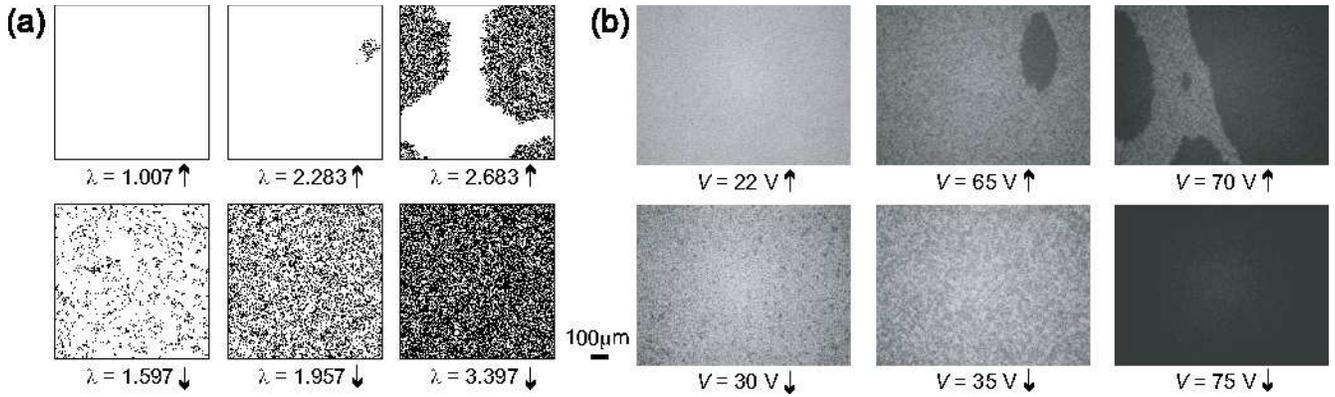}
  \caption{Hysteresis observed in simulations and experiments. (a) Hysteresis of (2+1)D CP with $h'=10^{-2}$. Black regions denote active sites. The control parameter $\lambda$ is increased and then decreased in the range of $1 \leq \lambda \leq 3.4$ at the ramp rate of $r=0.001\unit{MCS^{-1}}$. The critical point for the model without nucleation is $\lambda_{\rm c}=1.64877(3)$ \cite{Dickman-PRE1999}. The arrows after the values of $\lambda$ denote whether they are increasing or decreasing. (b) Hysteresis in the electrohydrodynamic convection, where the same cell as in Ref.\ \cite{Takeuchi_etal-PRL2007} is used. The control parameter, applied voltage $V$, is ramped in the range of $22\unit{V} \leq V \leq 75\unit{V}$ at the rate of $r=1.71\unit{V/s}$ with fixed frequency of $250\unit{Hz}$. The critical voltage is $V_{\rm c} \approx 35\unit{V}$ \cite{Takeuchi_etal-PRL2007}. Darker regions correspond to DSM2, the active state. Note that the global intensity and contrast are adjusted for the sake of clarity, and that DSM1 and DSM2 coexist in the two images at the lower left.}
  \label{fig:hysteresis}
 \end{center}
\end{figure*}%

\begin{figure}[t]
 \begin{center}
  \includegraphics[clip]{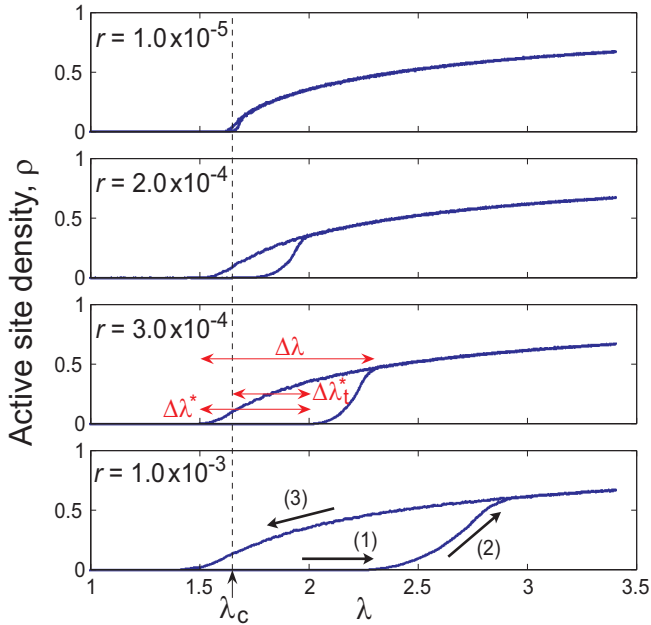}
  \caption{(Color online) Typical hysteresis loops for four different ramp rates $r$ in (2+1)D CP with $h'=10^{-2}$. Note that the ratios of the four values of $r$ are chosen to be approximately the same as in Fig.\ 1 of Ref.\ \cite{Kai_etal-PRL1990} to allow the comparison (see also Note \cite{Footnote1}). The hysteretic process can be decomposed into three stages as indicated in the bottom figure.}
  \label{fig:HysteresisLoops}
 \end{center}
\end{figure}%

The model behaves similarly to the turbulence of liquid crystals
 in many aspects. For instance $\lambda \gg \lambda_{\rm c}$
 and initial conditions of $s_{i,j}=0$ everywhere
 lead to a nucleus growth after sufficient time has passed,
 which faithfully reproduces experiments.
In particular, the model exhibits hysteresis
 as shown in Fig.\ \ref{fig:hysteresis}(a) and Movie S1 \cite{EPAPS}
 when $\lambda$ is increased from $\lambda<\lambda_{\rm c}$
 to $\lambda>\lambda_{\rm c}$ at a constant ramp rate $r$
 and then decreased at the same speed.
The hysteretic process can be decomposed into three stages
 as indicated in the bottom of Fig.\ \ref{fig:HysteresisLoops}.
Let us start from the uniformly inactive state and increase $\lambda$.
First, active clusters do not emerge 
 even for $\lambda>\lambda_{\rm c}$
 due to the very low nucleation rate (1st stage).
However, once a spontaneous nucleation occurs,
 the active nucleus grows and finally covers the whole system
 because of $\lambda>\lambda_{\rm c}$ (2nd stage).
The density of active sites, $\rho$, saturates
 at the steady state value $\rho_\text{steady}(\lambda)$.
On the other hand, when $\lambda$ is decreased,
 the number of active sites decreases gradually and homogeneously
 contrary to the growing process,
 approximately following $\rho_\text{steady}(\lambda)$
 (3rd stage).
This strikingly resembles what is observed in the liquid crystal experiments
 [Fig.\ \ref{fig:hysteresis}(b), Movie S2 \cite{EPAPS},
 Refs.\ \cite{Kai_etal-PRL1990,Footnote1}].
Note that the observed hysteresis
 both in the experiments and in the simulations
 is not a stationary property of the system,
 as would imply a first order transition,
 but rather a dynamical effect owing to the sweep of the parameter.

\begin{figure}[t]
 \begin{center}
  \includegraphics[clip]{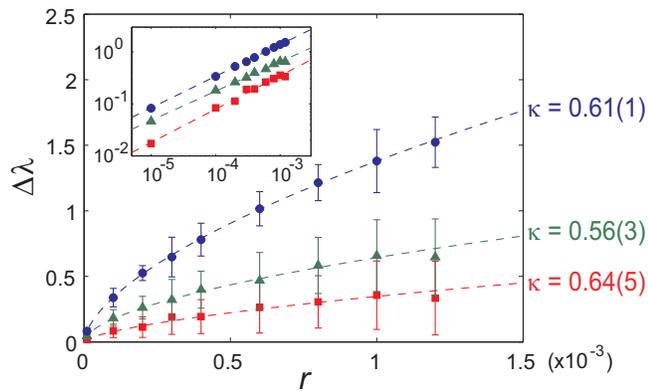}
  \caption{(Color online) Widths of the loops $\Delta\lambda$ (disk), $\Delta\lambda^*$ (triangle), and $\Delta\lambda^*_t$ (square) with respect to the ramp rate $r$, in the case of (2+1)D CP with $h'=10^{-2}$. The symbols and errorbars indicate means and standard deviations, respectively, of 50 independent runs. Dashed curves denote the results of the fitting to the power law $\Delta\lambda, \Delta\lambda^*, \Delta\lambda^*_t \sim r^\kappa$.
The inset shows the same data in logarithmic scales.}
  \label{fig:WidthVsRampRate}
 \end{center}
\end{figure}

The dependence on the ramp rate $r$ is shown
 in Fig.\ \ref{fig:HysteresisLoops},
 which is again very similar to the corresponding experiments
 \cite{Kai_etal-PRL1990,Footnote1}.
The widths of the hysteresis loops
 $\Delta\lambda$ and $\Delta\lambda^*$,
 defined as in Fig.\ \ref{fig:HysteresisLoops},
 clearly exhibit the power law dependence
 $\Delta\lambda, \Delta\lambda^* \sim r^\kappa$   
 [Fig.\ \ref{fig:WidthVsRampRate} (disks and triangles)],
 with $\kappa = 0.61(1)$ for $\Delta\lambda$ and
 $\kappa = 0.56(3)$ for $\Delta\lambda^*$.
Here the ranges of error correspond to 95\% confidence intervals
 in the sense of Student's t.
They are in good agreement with the experimental value
 $\kappa = 0.5$-$0.6$ \cite{Kai_etal-PRL1990,Kai-PC1}.

Besides the agreement between the simulations and the experiments,
 the exponent $\kappa$ can also be derived
 only by assuming DP criticality with a very low probability
 for spontaneous nucleation.
For absorbing phase transitions,
 the probability $P_\infty$ with which
 an active site survives forever
 grows algebraically as $P_\infty \sim \varepsilon^{\beta'}$
 for $\varepsilon \equiv \lambda-\lambda_{\rm c} > 0$,
 where $\beta'$ constitutes one of the critical exponents
 characterizing these transitions.
(Note that for the DP class
 the so-called ``rapidity'' symmetry implies $\beta' = \beta$
 \cite{Hinrichsen-AdvPhys2000,Grassberger_delaTorre-AnnPhys1979},
 where $\beta$ is the critical exponent
 corresponding to the stationary active site density $\rho_\text{steady}$.)
Suppose $\varepsilon$ is increased linearly
 as $\varepsilon(t) = rt$ and
 a nucleus appears and grows at time $t=T$,
 and assume that the ramp rate $r$ is so slow that
 the finite-time survival probability converges to $P_\infty$
 before the control parameter significantly changes,
 the following relation then approximately holds:
\begin{equation}
 1 \approx \int_0^T h'P_\infty (\varepsilon(t)) \mathrm{d}t \sim h'r^{\beta'} T^{\beta'+1},  \label{eq:RoughDerivationKappa}
\end{equation}
 and thus the width of the hysteresis is
\begin{equation}
 \Delta\lambda^*_t \equiv rT \sim r^{1/(\beta'+1)}.  \label{eq:Kappa}
\end{equation}
It gives the exponent for the hysteresis as
 $\kappa = 1/(\beta'+1) = 0.632(2)$ for the (2+1)D DP
 \cite{DPexponents}.
Of course the assumed nucleation process is stochastic,
 so that, strictly, one should deal with the average width
 $\langle \Delta\lambda^*_t \rangle$ based on the probabilistic distribution.
This more rigorous approach is also  straightforward.
With $P_0(t)$ being the probability that a nucleus
 does not appear and grow until time $t$,
 the probability that such a nucleation first occurs
 between time $t$ and $t+\mathrm{d}t$ is written as
\begin{align}
 -\mathrm{d}P_0(t) &= P_0(t) \cdot h'P_\infty(\varepsilon(t)) \mathrm{d} t \notag \\
 &= Ch'r^{\beta'} t^{\beta'} \exp \left( -\frac{Ch'r^{\beta'}}{\beta'+1}t^{\beta'+1} \right) \mathrm{d} t,
\end{align}
 where $C$ is defined by $P_\infty = C\varepsilon^{\beta'}$.
This gives the average of the hysteresis width as
\begin{align}
 \langle \Delta\lambda^*_t \rangle &= r \int_0^\infty t \left(-\frac{\mathrm{d}P_0(t)}{\mathrm{d}t}\right) \mathrm{d}t \notag \\
 &= \Gamma \left(\frac{\beta'+2}{\beta'+1}\right) \left[ \frac{(\beta'+1)r}{Ch'}\right]^{1/(\beta'+1)},
\end{align}
 which confirms Eq.\ \eqref{eq:Kappa}.
Note that the standard deviation also obeys the same power law
 (with a different coefficient),
 since the stochastic process at play is essentially Poissonian.

The derived value of $\kappa = 0.632(2)$
 is slightly larger than the numerical and experimental values.
This stems from the use of different definitions
 for the lower bound of the loop:
 the merging point of the two curves defining the loop is used
 for the experiments and simulations
 ($\Delta\lambda$ and $\Delta\lambda^*$),
 whereas, theoretically, the exact critical point $\lambda_c$ is used to
 define the lower bound.
Adopting the latter definition for the simulations
 ($\Delta\lambda^*_t$ in Fig.\ \ref{fig:HysteresisLoops}),
$\kappa = 0.64(5)$ is obtained (Fig.\ \ref{fig:WidthVsRampRate}),
 which is now in close agreement with the theoretical value.
Thus the picture based on DP with a weakly broken absorbing state
 quantitatively explains the observed hysteresis.
The above derivation also indicates that
 the scaling of hysteresis loops is seen for such values of $h'$ and $r$ that
 nucleations, including those with a short lifetime,
 occur several times within the range where
 the scaling $P_\infty \sim \varepsilon^{\beta'}$ holds.
 
\begin{table}[t]
 \begin{center}
  \caption{Hysteresis exponent $\kappa$ for several models.}
  \label{tbl:WidthsVsRampRateUniversality}
  \catcode`?=\active \def?{\phantom{0}}
  \begin{tabular}{lllll} \hline\hline
   & \multicolumn{3}{c}{Exponent $\kappa$ for} & \\ \cline{2-4}
   \multicolumn{1}{c}{Model\footnotemark[1]} & \multicolumn{1}{c}{$\Delta\lambda$} & \multicolumn{1}{c}{$\Delta\lambda^*$} & \multicolumn{1}{c}{$\Delta\lambda^*_t$} & \multicolumn{1}{c}{$1/(\beta'+1)$} \\ \hline
   (2+1)D CP (PCA) & $0.61(1)$ & $0.56(3)$ & $0.64(5)$ & $0.632(2)$\footnotemark[2] \\
   (2+1)D CP & $0.61(2)$ & $0.61(4)$ & $0.65(7)$ & $0.632(2)$\footnotemark[2] \\
   (1+1)D CP (PCA) & $0.69(1)$ & $0.73(3)$ & $0.81(4)$ & $0.783$\footnotemark[2] \\
   (1+1)D site DP\footnotemark[3] & $0.68(2)$ & $0.71(5)$ & $0.82(7)$ & $0.783$\footnotemark[2] \\
   (2+1)D voter-like\footnotemark[4]\footnotemark[5] & $0.465(14)$ & $0.460(17)$ & $0.47(4)$ & $0.5$ \\ \hline
  \footnotetext[1]{System sizes and nucleation rates are set to $L=4096$ and $h'=10^{-4}$ for (1+1)D, and to $L=256$ and $h'=10^{-2}$ for (2+1)D.}
  \footnotetext[2]{Values of the DP exponent $\beta'$ are from Ref.\ \cite{Jensen-JPhysA1999} for (1+1)D and from Refs.\ \cite{DPexponents} for (2+1)D.}
  \footnotetext[3]{Simulations are performed in the Domany-Kinzel lattice.}
  \footnotetext[4]{Kinetic Ising model with spin-flip probability $p_{sH}$ is considered, where $s = \pm 1$ and $H \in \{-4, -2, 0, 2, 4\}$ denote a spin and its local field, respectively. $p_{-2}$ is swept here with the other parameters fixed at $p_4 = h = 10^{-2}/L^2, p_2 = 0.17, p_0 = 0.5, p_{-4} = 0.68$. This model shows a transition in the voter universality class \cite{Dornic_etal-PRL2001}.}
  \footnotetext[5]{Hysteresis is measured in terms of the density of interfaces (i.e., the fraction of $+-$ pairs) instead of the active site density (i.e., magnetization), since the former characterizes the voter class better \cite{Dornic_etal-PRL2001} and shows faster relaxation.}
  \end{tabular}
 \end{center}
\end{table}

\begin{figure}[t]
 \begin{center}
  \includegraphics[clip]{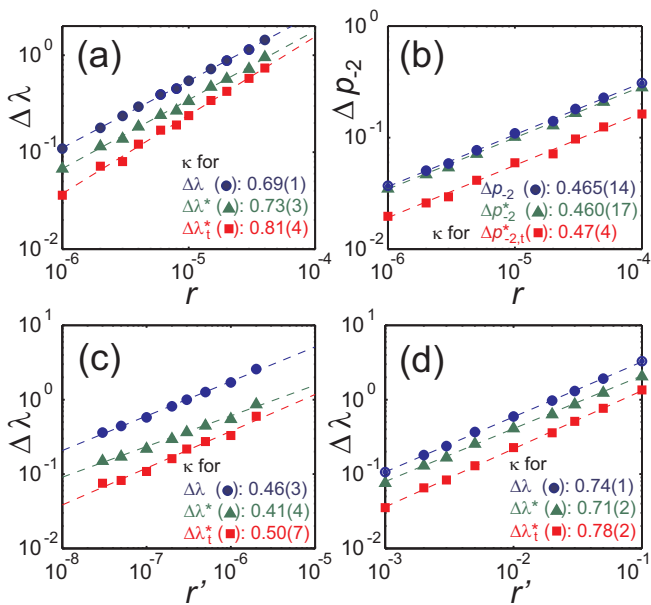}
  \caption{(Color online) Hysteresis scaling for (1+1)D CP (a), (2+1)D voter-like(b), (2+1)D CP with quadratic ramping $\varepsilon(t) = r't^2$ (c), and (2+1)D CP with square root ramping $\varepsilon(t) = r't^{1/2}$ (d). The estimates of the exponent $\kappa$ for $\Delta\lambda^*_t$ agree with the theoretical values, namely, $0.783$ (a), $0.5$ (b), $0.462(2)$ (c), and $0.774(2)$ (d).}
  \label{fig:VariousWidthVsRampRate}
 \end{center}
\end{figure}%

Given that the only assumption was
 criticality of an absorbing transition
 together with very rare spontaneous nucleations,
the observed scaling of hysteresis
 with the exponent $\kappa = 1/(\beta'+1)$ is expected to be found universally
 in systems that exhibit
 quasi-absorbing transitions.
This is confirmed by performing simulations in different dimensions
 and for different models and universality classes
 [Table \ref{tbl:WidthsVsRampRateUniversality}
 and Fig.\ \ref{fig:VariousWidthVsRampRate} (a)(b)].
In all cases, the measured $\kappa$ values for $\Delta\lambda^*_t$
 are in good agreement with those derived by Eq.\ \eqref{eq:Kappa}.
The loop scaling is also robust to situations when
 the ramp rate $r$ and/or the nucleation probability $h'$
 vary with time or the control parameter.
As long as they are nonzero and analytic at criticality,
 this gives only higher order corrections
 to Eq.\ \eqref{eq:RoughDerivationKappa}
 and does not affect the final result when $r \to 0$.
Even if this condition is not satisfied,
 the corrected form of Eq.\ \eqref{eq:Kappa} can be calculated, for example 
 in the case of nonlinear ramping $\varepsilon(t) = r't^a$;
 the hysteresis exponent becomes then $\kappa = 1/(a\beta' + 1)$,
 which is numerically confirmed
 [Fig.\ \ref{fig:VariousWidthVsRampRate} (c)(d)].

Some experimental systems expected to belong to the DP class
 seem to lack strictly absorbing states due to residual nucleations
 \cite{Hinrichsen-AdvPhys2000,Rupp_etal-PRE2003}.
This suggests that the same hysteresis may be observed in such systems,
 for example
 with different alignments or at other transitions
 in the electrohydrodynamic convection
 \cite{Lucchetta_etal-PRE1999,Oikawa_etal-PTPS2006}.
A much more intriguing candidate can be found
 in the field of quantum turbulence \cite{QuantumTurbulence}.
Recently, a number of experimental studies
 on transitions to turbulence in superfluid \elem{4}{He}
 have reported hysteresis
 \cite{Jager_etal-PRL1995,Bradley_etal-JLTP2005,Hashimoto_etal-JLTP2007,Hanninen_etal-PRB2007}
 and temporal intermittency in local state of turbulence
 \cite{Niemetz_Schoepe-JLTP2004,Bradley_etal-JLTP2005,Hashimoto_etal-JLTP2007,Hanninen_etal-PRB2007}.
The existence of a (quasi-)absorbing state is also expected
 due to the quantum topological constraint.
All of these facts suggest that
 an absorbing transition to STI may take place in this superfluid system.
Although it seems technically difficult to examine
 conventional critical phenomena of absorbing transitions directly there,
 scaling of hysteresis loops may be more easily accessible and would
allow to decide about the corresponding universality class.

In conclusion, the hysteresis loop scaling
 experimentally observed before
 at the DSM1-DSM2 transition of liquid cristal convection was explained
 by assuming DP dynamics with very rare spontaneous nucleations.
This implies that DSM1 is probably only quasi-absorbing
 in the liquid crystal system.
Moreover, scaling of hysteresis loops
 $\Delta\lambda \sim r^\kappa$ with $\kappa = 1/(\beta'+1)$
was demonstrated to be able to decide the universality class
of transitions into a quasi-absorbing state.
These results may also be used to analyze critical phenomena in systems
 where measurable quantities are so limited
 that usual approaches to absorbing phase transitions cannot be adopted,
 such as in superfluid turbulence.

I am grateful to M. Sano, H. Chat\'e, I. Dornic, S. Kai, M. Kobayashi,
 M. Kuroda, N. Oikawa, H. Park, M. Tsubota, and H. Yano
 for fruitful discussions.
This work is partly supported
 by JSPS Research Fellowships
 for Young Scientists.



\begin{thebibliography}{99}

\bibitem{Hinrichsen-AdvPhys2000}
 H. Hinrichsen, Adv. Phys. \textbf{49}, 815 (2000).

\bibitem{Janssen-ZPhysB1981}
 H. K. Janssen, Z. Phys. B \textbf{42}, 151 (1981).

\bibitem{Grassberger-ZPhysB1982}
 P. Grassberger, Z. Phys. B \textbf{47}, 365 (1982).

\bibitem{Takeuchi_etal-PRL2007}
 K. A. Takeuchi, M. Kuroda, H. Chat\'e, and M. Sano, Phys. Rev. Lett. \textbf{99}, 234503 (2007).

\bibitem{Pomeau-PhysD1986}
 Y. Pomeau, Physica D \textbf{23}, 3 (1986).

\bibitem{Kai_etal-PRL1990}
 S. Kai, W. Zimmermann, M. Andoh, and N. Chizumi, Phys. Rev. Lett. \textbf{64}, 1111 (1990).

\bibitem{Kai-PC1}
 S. Kai (private communication).

\bibitem{Harris-AnnProb1974}
 T. E. Harris, Ann. Prob. \textbf{2}, 969 (1974).

\bibitem{Dickman-PRE1999}
 R. Dickman, Phys. Rev. E \textbf{60}, R2441 (1999).

\bibitem{Luebeck-IJMB2004}
 S. L\"ubeck, Int. J. Mod. Phys. B \textbf{18}, 3977 (2004).

\bibitem{EPAPS}
 See EPAPS Document No. [number will be inserted by publisher] for Movies S1 and S2. For more information on EPAPS, see http://www.aip.org/pubservs/epaps.html.

\bibitem{Footnote1}
 Hysteresis loops in Fig \ref{fig:HysteresisLoops} should be compared with Fig.\ 1 in Ref.\ \cite{Kai_etal-PRL1990}. The global light transmittance is plotted in the latter, which can be roughly translated into $\rho$ by turning the figure upside down.

\bibitem{Grassberger_delaTorre-AnnPhys1979}
 P. Grassberger and A. de la Torre, Ann. Phys. \textbf{122}, 373 (1979).

\bibitem{DPexponents}
 P. Grassberger and Y. C. Zhang, Physica A \textbf{224}, 169 (1996);
 C. A. Voigt and R. M. Ziff, Phys. Rev. E \textbf{56}, R6241 (1997).

\bibitem{Jensen-JPhysA1999}
 I. Jensen, J. Phys. A: Math. Gen. \textbf{32}, 5233 (1999).




\bibitem{Dornic_etal-PRL2001}
 I. Dornic, H. Chat\'e, J. Chave, and H. Hinrichsen, Phys. Rev. Lett. \textbf{87}, 045701 (2001).

\bibitem{Rupp_etal-PRE2003}
 P. Rupp, R. Richter, and I. Rehberg, Phys. Rev. E \textbf{67}, 036209 (2003).

\bibitem{Lucchetta_etal-PRE1999}
 D. E. Lucchetta, N. Scaramuzza, G. Strangi, and C. Versace, Phys. Rev. E \textbf{60}, 610 (1999).

\bibitem{Oikawa_etal-PTPS2006}
 N. Oikawa, Y. Hidaka, and S. Kai, Prog. Theor. Phys. Suppl. \textbf{161}, 320 (2006).

\bibitem{QuantumTurbulence}
 W. F. Vinen and R. J. Donnelly, Phys. Today \textbf{60}(4), 43 (2007); W. F. Vinen and J. J. Niemela, J. Low Temp. Phys. \textbf{128}, 167 (2002).

\bibitem{Jager_etal-PRL1995}
 J. J\"{a}ger, B. Schuderer, and W. Schoepe, Phys. Rev. Lett. \textbf{74}, 566 (1995).

\bibitem{Bradley_etal-JLTP2005}
 D. I. Bradley \textit{et al.}, J. Low Temp. Phys. \textbf{138}, 493 (2005).

\bibitem{Hashimoto_etal-JLTP2007}
 N. Hashimoto \textit{et al.}, J. Low Temp. Phys. \textbf{148}, 299 (2007).

\bibitem{Hanninen_etal-PRB2007}
 For a brief review, see R. H\"{a}nninen, M. Tsubota, and W. F. Vinen, Phys. Rev. B \textbf{75}, 064502 (2007).

\bibitem{Niemetz_Schoepe-JLTP2004}
 M. Niemetz and W. Schoepe, J. Low Temp. Phys. \textbf{135}, 447 (2004).

\end{thebibliography}
\end{document}